\documentclass[final,5p,times,twocolumn]{elsarticle}

\usepackage{graphicx}
\usepackage{dcolumn}
\usepackage{bm}
\usepackage{color}
\usepackage{amssymb}
\usepackage{amsthm}
\usepackage{mathtools}

\usepackage{setspace}
\usepackage{amsmath}
\usepackage{hyperref}
\usepackage{ulem}
\usepackage{comment}
\usepackage [latin1]{inputenc}

\newcommand{\be}{\begin{equation}}
\newcommand{\ee}{\end{equation}}
\newcommand{\ba}{\begin{eqnarray}}
\newcommand{\ea}{\end{eqnarray}}
\newcommand{\bd}{\begin{displaymath}}
\newcommand{\ed}{\end{displaymath}}
\newcommand{\bea}{\begin{eqnarray}}
\newcommand{\eea}{\end{eqnarray}}

\newcommand{\beq}{\begin{equation}}
\newcommand{\beqar}{\begin{eqnarray}}
\newcommand{\eeq}[1]{\label{#1} \end{equation}}
\newcommand{\eeqar}[1]{\label{#1} \end{eqnarray}}

\journal{Nanophotonics}

\begin{document}

\begin{frontmatter}

\title{Kinetic Model Evaluation of
Dynamical Properties of Nanaorod Antennas Embedded in
a Polymer Carrying the Nuclei of Fusion Fuel
\\}

\author{
Istv\'an Papp $^{1,2}$,
Larissa Bravina $^4$,
M\'aria Csete  $^{1,5}$,
Archana Kumari $^{1,2}$,
Igor N. Mishustin  $^{6}$, \\
Anton Motornenko   $^6$, 
P\'eter R\'acz $^{1,2}$,
Leonid M. Satarov  $^{6}$,
Horst St\"ocker  $^{6,7,8}$,
Daniel D. Strottman $^{9}$,\\
Andr\'as Szenes  $^{1,5}$,
D\'avid Vass   $^{1,5}$,
\'Agnes Nagyn\'e Szokol $^{1,2}$,
Judit K\'am\'an $^{1,2}$,
Attila Bony\'ar $^{10}$,\\
Tam\'as S. Bir\'o $^{1,2}$,
L\'aszl\'o P. Csernai $^{1,2,3,6,11}$,
Norbert Kro\'o $^{1,2,12}$\\
(part of NAPLIFE Collaboration)\\}
\bigskip

\address{
$^{1}$ Wigner Research Centre for Physics, Budapest, Hungary\\
$^2$ Hungarian Bureau for Research Development and Innovation \\
$^3$ Dept. of Physics and Technology, University of Bergen, Norway\\
$^4$ Department of Physics, University of Oslo, Norway\\
$^5$ Dept. of Optics and Quantum Electronics, Univ. of Szeged, Hungary\\
$^6$ Frankfurt Institute for Advanced Studies, Frankfurt/Main, Germany\\
$^7$ Inst. f\"ur Theoretische Physik, Goethe Universit\"at, Frankfurt/Main, Germany\\
$^{8}$ GSI Helmholtzzentrum f\"ur Schwerionenforschung GmbH, Darmstadt, Germany\\
$^{9}$ Los Alamos National Laboratory, Los Alamos, New Mexico, USA \\
$^{10}$ Department of Electronics Technology,  Faculty of Electrical Engineering and Informatics\\
Budapest University of Technology and Economics, Hungary\\
$^{11}$ Csernai Consult Bergen, Bergen, Norway \\
$^{12}$ Hungarian Academy of Sciences, Budapest, Hungary\\[-2em]
}

\begin{abstract}{
Recently laser induced fusion with simultaneous volume ignition, 
a spin-off from relativistic heavy ion collisions,
was proposed, where implanted nanoantennas regulated and amplified the 
light absorption in the fusion target. Studies of resilience of the 
nanoantennas was published recently in vacuum. These studies are
extended to nanoantennas embedded into a polymer, which modifies the 
nanoantenna's lifetime and absorption properties.
}
\end{abstract}

\end{frontmatter}

\section{Introduction}

%
NAnoPlasmonic Laser Inertial Fusion Experiments (NAPLIFE)  
\cite{CsEA2020} 
is an improved way to 
achieve laser driven fusion in a non-thermal, collider configuration
to avoid instabilities during ignition.
It is based on simultaneous (or "time-like") ignition 
\cite{Cs1987,CS2015},
with enhanced energy absorption 
with the help of nanoantennas implanted into the target material
\cite{CsKP2018}. 
This should prevent the development of 
the mechanical Rayleigh-Taylor 
instability. Furthermore, the nuclear burning should not propagate
from a central hot spot to the outside edge as the ignition is
simultaneous in the whole volume.

Non-equilibrium and linear colliding configuration 
have been introduced already 
\cite{Barbarino2015,Bonasera2019}.
Here we study the idea of layered flat target fuel with
embedded nanorod antennas, that regulate laser light 
absorption to enforce simultaneous ignition.
We plan a seven-layer flat target with different
nanorod densities. 
\cite{CseteX2022,BA2020,OMEX}.
In order to prepare such a layered target, the ignition fuel
(e.g. deuterium, D, tritium, T, or other nuclei for fusion)
are embedded into a hard thin polymer material of seven,
3$\mu$m thick layers. These 
 polymers are
Urethane Dimethacrylate  (UDMA) and 
Triethylene glycol dimethacrylate (TEGDMA) in (3:1) mass ratio 
\cite{UDMA1,BA2022}.
The UDMA polymer molecule contains 470 nuclei, 38 of them are Hydrogen.
One can also use deuterized UDMA, where some of the Hydrogens are
replaced by Deuterium atoms.

\begin{figure}[h]  
\begin{center}
\includegraphics[width=0.99\columnwidth]{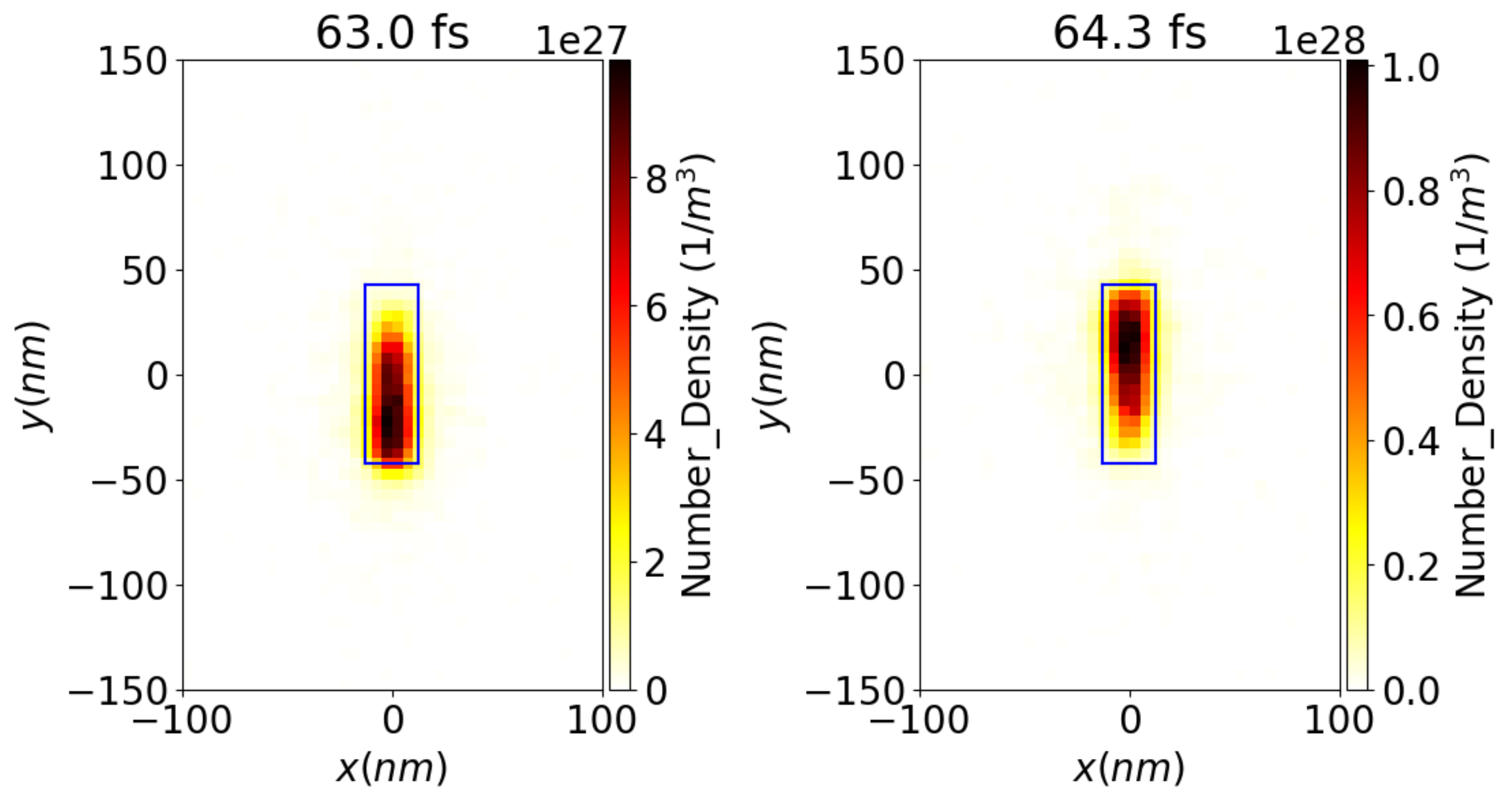}
\end{center}
\vskip -0.5cm
\caption{(color online) Cross section of 
the 25 nm (diameter) x 85 nm nanorod showing
Number density of electrons at the tips at different times $t$, 
half of the light wave time period apart, show electrons leaving 
the nanorod. The number of electron 
marker-particles inside the simulation box will decrease 
with a significant amount by the end
of simulation at 120 fs.}
\label{Fig-1}
\end{figure}

\section{Dynamics of the resonance in the nanorod}

When a resonant nanoparticle of a size related to the effective 
wavelength of light a localised surface plasmon (LSP) is created. 
When the coherently oscillating electric field irradiates the metallic 
nanoparticle it causes the conduction electrons to oscillate also. 
The Coulomb attraction between electrons and nuclei produces a 
restoring force when the electron cloud is moved from its initial location.
The electron cloud oscillates due to this force. The effective 
electron mass, the size and form of the charge distribution, 
and the electron density all contribute to the oscillation frequency.
The LSP has two key effects: it dramatically increases electric fields 
near the nanoparticle's surface and it increases optical 
absorption at the plasmon resonance frequency. 
The form of the nanoparticle can also be used to adjust surface plasmon 
resonance
\cite{Kelly2003,Maier2007}.

A recent kinetic theoretical study analyzed the resilience
of nanorod antennas under a short laser pulse irradiation
in vacuum
\cite{PRX-E2022}.

Now we extend these studies 
to nanorod antennas embedded into the UDMA polymer.
Here we consider the refracting index of UDMA ($n$=1.53) \cite{BA2022}, which
slows down the propagation of light. The short laser pulse is
chosen to have a length needed to propagate across the target of
$7\cdot 3\mu{\rm m} = 21\mu{\rm m}$ thickness. The nanorods are 
orthogonal to the direction of laser irradiation in this model study.

\begin{figure}  
\begin{center}
\includegraphics[width=0.99\columnwidth]{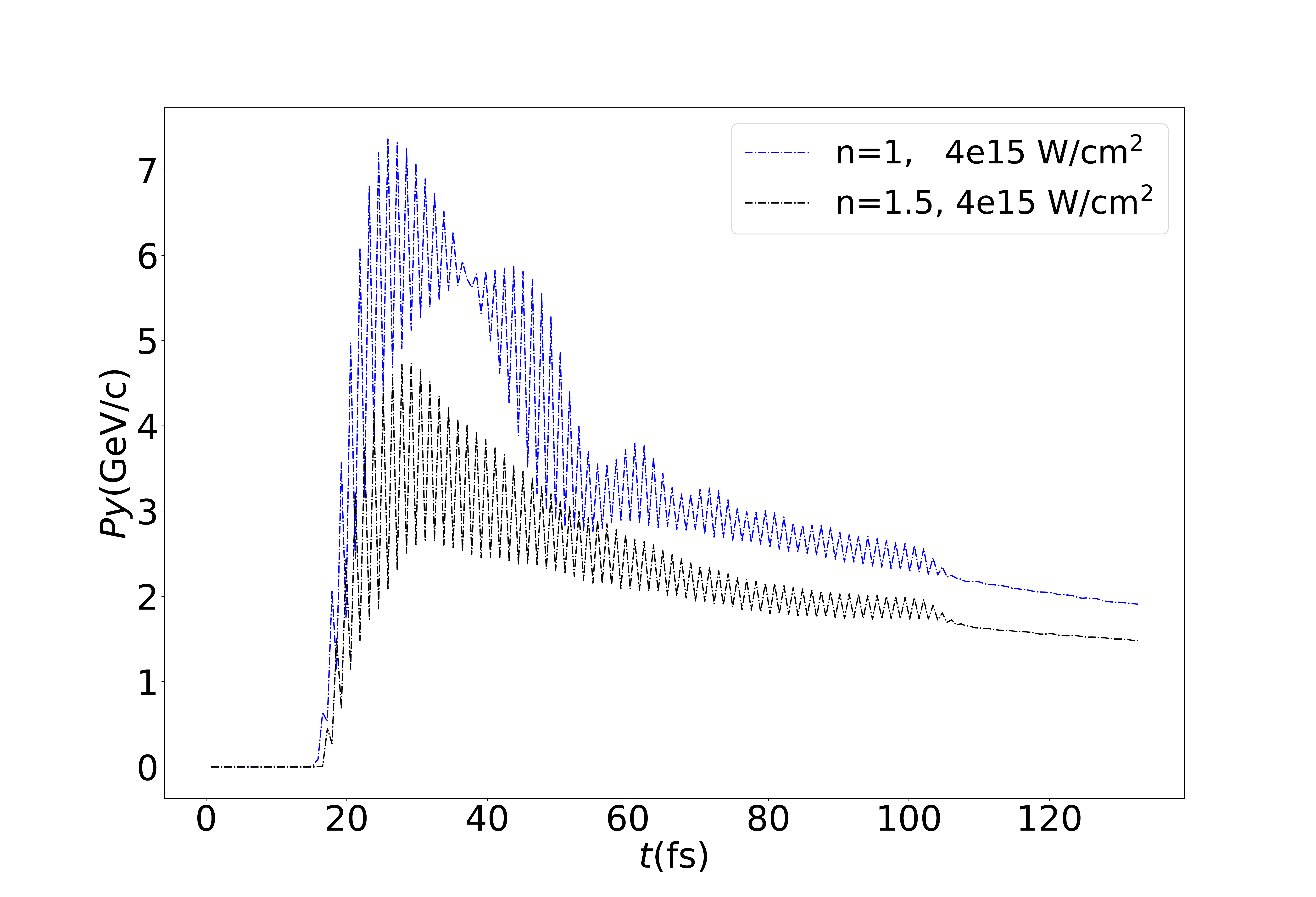}
\end{center}
\vskip -0.5cm
\caption{(color online)
We consider a laser pulse of intensity 
$I= 4 \cdot 10^{15}$W/cm$^2$ and duration of $106$fs.
Here we show the time dependence of the total $y$ directed 
momentum of the conducting electrons in the nanorod. 
The nanorod is in surrounding UDMA polymer (black line) 
and in vacuum (blue line).
The UDMA polymer decreases the momentum of the emitted electrons 
considerably compared to the emission to vacuum. An interesting 
"beat" interference is occurring in case of emission to vacuum. 
Apparently the nanorod antenna in vacuum has two slightly different 
resonant frequencies. The rate of beat is 
$\sim 5\cdot 10^{13}$Hz, 
while the frequency of the driving laser beam is $3.8 \cdot 10^{14}$Hz
($T=2.65$ fs).
In UDMA the beating does not occur as the somewhat random surface 
between the gold nanorod and the kinetic model representation of UDMA
does not lead to two distinct resonance frequencies.
}
\vskip -0.8cm
\label{Fig-2}
\end{figure}

The conduction electrons show behavior of strongly coupled plasma 
\cite{Nov2007}. 
The gold antennas are smaller than the half wavelength of the irradiated light. At optical frequencies the classical ideal half-wavelength 
dipole antenna scaling of rod with length $L = \lambda/2 $  breaks down.
Here instead an effective wavelength needs to be considered 
\cite{Nov2007}.
When the nanorods are embedded in a surrounding medium different 
from vacuum the effective wavelength scales as follows:
\begin{equation}
\begin{aligned}
\label{eq1}
\frac{\lambda_{eff}}{2R\pi} =
13.74 - 0.12 [ \varepsilon_\infty+\varepsilon_s141.04 ]/\varepsilon_s \\ 
- \frac{2}{\pi} +  \frac{\lambda}{\lambda_p} 0.12 
\sqrt{\varepsilon_\infty+\varepsilon_s141.04}/\varepsilon_s
\end{aligned}
\end{equation}
where 
$ \varepsilon_\infty = 11$ 
is the dielectric function in the infinite frequency limit 
\cite{eps-inf} 
and $\lambda_p$= 138 nm is the plasma wavelength for gold. 
The propagation velocity of light inside the medium is reduced to 
$c_s=1/\sqrt{\varepsilon_s}$, 
therefore $\varepsilon_s = n^2$. 
This is a good motivation for using Particle in Cell (PIC) methods for 
studying this behavior. We used similar principles as described in 
\cite{PRX-E2022}, 
using the EPOCH computing package 
\cite{Arber2015,Nanbu1998,Perez2012}. 
We considered nanorod antennas with partly ionised gold atoms 
with three conducting electrons per gold atoms.

Initially electrons in the $\lambda_{eff}/2 = 85$nm \cite{BA2020}
nanorod antenna follow the phase of the laser irradiation
with $t = $2.65 fs period. 
With time electrons diffuse out of the nanorod,
mainly at its two ends (Fig. \ref{Fig-1}). The potential wall to UDMA
keeping the electrons in the nanorod is apparently smaller than 
in the case of surrounding vacuum
\cite{PRX-E2022}.

Simulation studies using the COMSOL Multiphysics
Finite Eement Method (FEM) package
with many parameters 
found absorptivities between 0.085 and 0.192 for nanorod antennas 
\cite{CseteX2022}. 
By varying the density of implanted nanoantennas one could achieve 
almost uniform integrated energy absorption at a given overlapping time 
of 240 fs for two counter-propagating 120 fs laser pulses 
\cite{CseteX2022,P-EA2021}.

The beam intensity utilized was $I=4\cdot10^{15}$W/cm$^2$ so that the 
plasmonic nanoantennas are not destroyed before the laser pulse passes.
This damage threshold also depends on the geometry and size of the 
nanoantennas. 

We compared the time dependence of  momentum of escaping electrons 
in vacuum and in UDMA. The magnitude and the dynamics of electron 
emission is quite different, as shown in Fig. \ref{Fig-2}. 

The gold nanorods may lose a few 
electrons from the conduction band without destroying their 
solid structure, so they still 
can enhance the absorptivity of the ignition pulse for
about 20-50 fs, leaving enough time for the full ignition. 

For the emission of a single electron from gold plasmonic nanoantennas
four 795 nm photons are needed. On the other hand the incoming pulse 
generates a surface plasmon, which may later emit electrons. 
This indirect process is more frequent than the direct emission by four 
photons
\cite{MP2000,Farkas}.

Similarly to the analysis in ref.
\cite{PRX-E2022} 
we now study the energy transfer dynamics from the laser irradiation
to the target, with and without nanorods.

\begin{figure*}[!h]  
\begin{center}
\resizebox{0.99\textwidth}{!}{
\includegraphics[width=0.99\columnwidth]{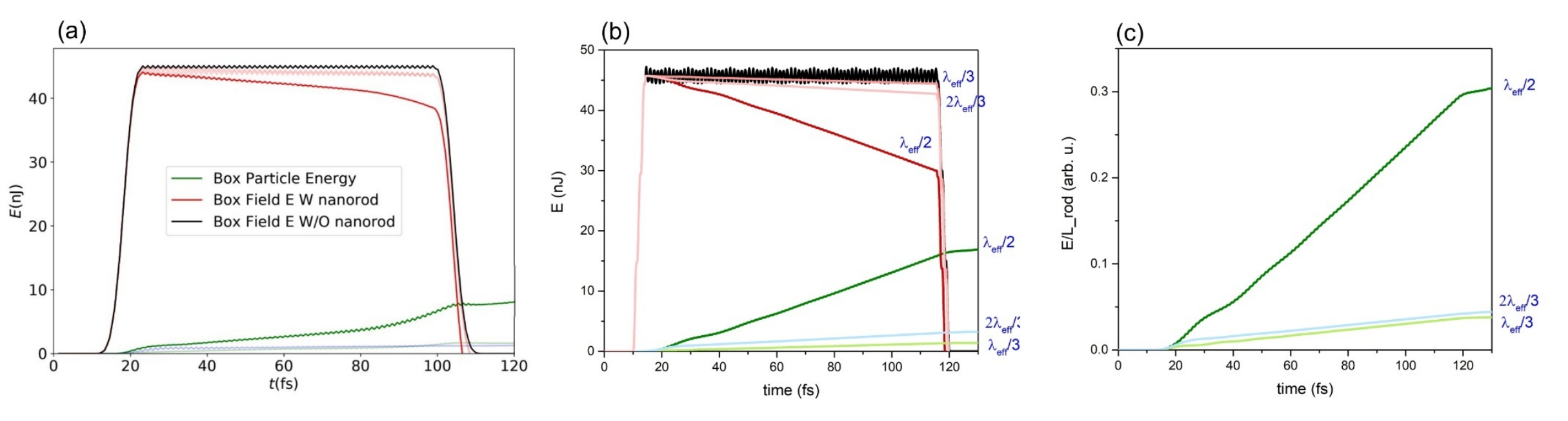}}
\end{center}
\vskip -0.5cm
\caption{(color online)
Optical response of the gold nanorod with different numerical methods
and lengths, $L=\lambda_{eff}/2$, $\lambda_{eff}/3$ and $2\lambda_{eff}/3$,
(a) PIC, (b) FEM and (c) FEM with normalized values
to unit antenna length.
The tendencies of the time-evolution of the nanorod energy determined by 
PIC and FEM are in very good agreement. 
The energy in the calculation box increases rapidly till about 20 fs, then 
it becomes constant (without nanoantenna) until the laser pulse lasts, at 
this moment the energy in the box drops to zero. 
With the nanoantenna in the box the resonant antenna absorbs a good part 
of the laser energy, (green line), while much less for the non resonant 
length antennas.
There is a quantitative difference between the rates of energy increase, 
namely the slope is significantly smaller for PIC computations, accordingly 
the value achieved at 106 fs is also smaller. The smaller slope is caused 
by the tunneling of the electrons out of the antenna that is included into 
the PIC computations but not in FEM.
The (c) figure shows that the difference between the two non-resonant 
antennas is caused by the antenna lengths, when this is removed by the 
length normalization the difference vanishes. }
\label{Fig-3}
\end{figure*}

Consider now an intense laser beam 
($\lambda$ = 795 nm in vacuum and 795/1.53 in UDMA),
with intensity $I = 4 \cdot 10^{15} $W/cm$^2$, irradiating a calculation
box (CB) of cross section 
$S_{CB} = 530 \cdot  530$nm$^2$ = $2.81 \cdot 10^{-9}$ cm$^2$  
and of length
$L_{CB} = \lambda =795$ nm,
with a step-function time profile of pulse length 
$T_P = 106$ fs
($\sim 40\lambda /c$). 
The laser pulse energy fraction falling into this box is 
$E_P = 1.19 \mu$J. 
In the geometrical middle we insert a single
nanorod antenna of length 85 nm and diameter 25 nm.
As the calculation box size ($\lambda$) is 1.53$\cdot$1/40th of the 
irradiation
pulse length (40$\lambda$), the initial and final transients are
negligible. See Fig. \ref{Fig-3}.

We used two different marker particle species, $42 500$ positively charged 
gold ions (+3) and three ($127 500$) conducting electrons for each, being 
careful to the neutral charge of the nanoantenna.  We define the size of 
the nanorod, indicating the limits where the particle number density 
becomes zero. The borders of which can be seen in Fig. \ref{Fig-1}.

We consider three situations, the box contains 
(i) vacuum,
(ii) UDMA polymer  and 
(iii) UDMA polymer with gold nanorod antenna 
in the middle of the box.

\begin{figure}[!h]  
\begin{center}
\includegraphics[width=0.99\columnwidth]{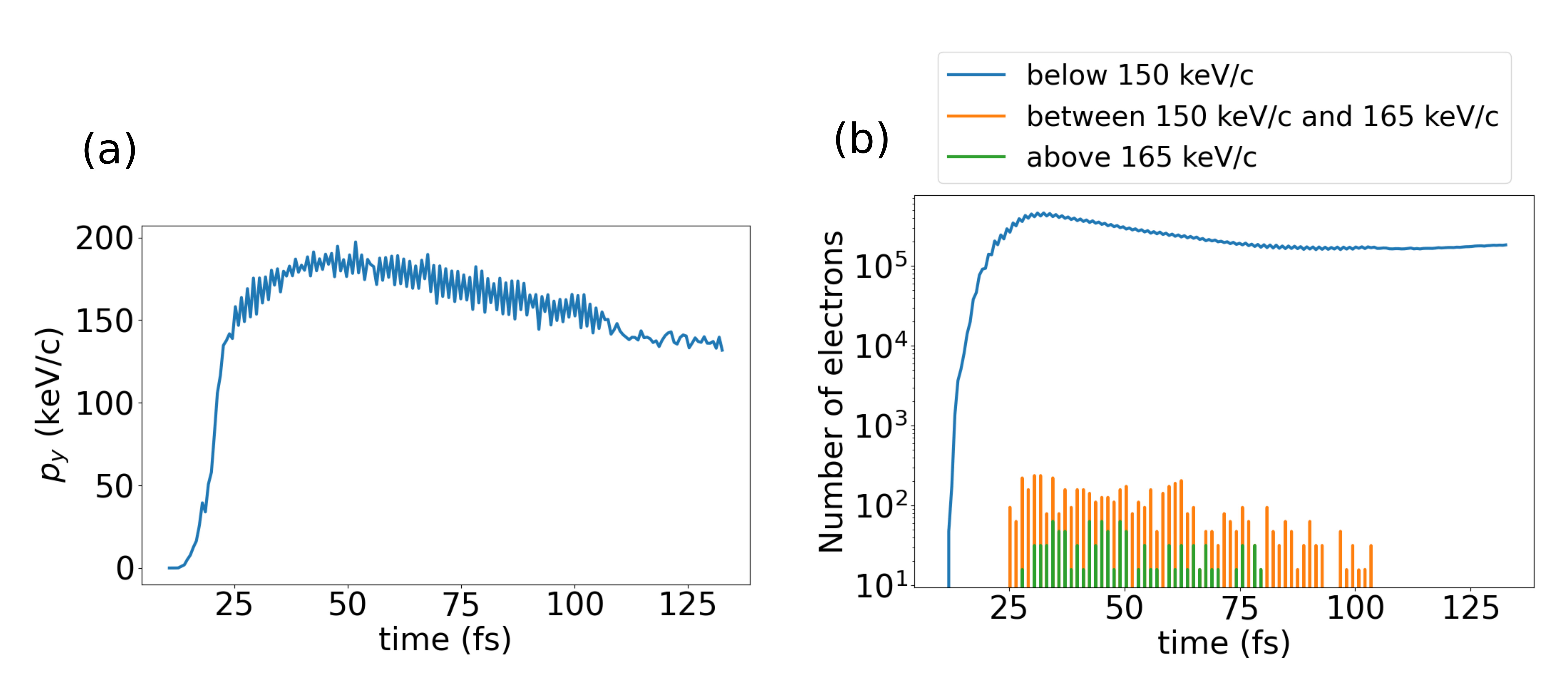}
\end{center}
\vskip -0.5cm
\caption{(color online) The behaviour of electrons leaving the nanorod. 
Flat plateau laser light reaches the nanorod with maximum 
intensity at around 20 fs and leaves the calculation box at 106 fs. 
Figure (a) indicates the maximum momentum in time reached by 
a spilled out electron in the $y$ direction. Figure (b) shows 
the distribution of electrons at different $y$ direction momentum 
values. Below 150 keV/c (blue line ) , above 150 keV/c 
(orange line) and above 165 kev/c (green line).
}
\label{F-4}
\end{figure}

We consider the following processes for direct absorption
to the UDMA. As UDMA is transparent the light 
absorption is minimal, while the refractive index is $n=1.53$. The absorption coefficient of the bare polymer matrix at this wavelength would be $\approx$ 0.3 cm$^{-1}$ while doped with gold nanorods it would reach 18 cm$^{-1}$ \cite{BA2022}.

In the EPOCH PIC kinetic 
plasma simulations model one is usually interested in charged particles, 
where the surrounding medium is vacuum. 
However, here we simulate metal nanoantenna
with conducting electrons approached as plasma and the UDMA polymer is 
taken into account 
with a relative electric permittivity different from vacuum. The wavelength 
inside the simulation box containing UDMA is also shrunk according to 
the refractive index.

\section{Conclusions and Outlook}

The result of this simulation shows that the resilience of the 
nanoantenna in the UDMA polymer is similar to the vacuum case. 
In case of vacuum  at 19 fs, when the maximum intensity of the laser
reaches the nanorod, most electrons, $N_e$=10$^3$, have 0.015 MeV/c momentum 
in the $y$ direction.
However at the time when the irradiation finishes at 
106 fs, around $N_e$=2 x 10$^2$ electrons remain at this momentum 
\cite{PRX-E2022}. 
Other electrons escape at the tip of the nanorod. At 43 fs the the number 
of leaving electrons reach the maximum while also achieving the maximum 
momentum in the $y$ direction (Fig. \ref{F-4}). Potential difference 
becomes $E_y = +/- 2.9 \times 10^{12}$ V/m $= +/- 2.9 x 10^3 $ V/nm. 
Maximum momentum of leaving electrons in the $y$ direction reaches 
0.3025 MeV/c in vacuum, in UDMA at the same time the maximum is lower 
at 0.1799 MeV/c.
The total momentum amplitude at this time is also lower in 
4.5 GeV/c in UDMA compared to 7.5 GeV/c in vacuum.

The decay time of the nanoantenna is somewhat shorter but sufficient
for the initial volume ignition. The absorbed energy from the CB to the 
nanoantenna and the polymer is slightly less, but this is understandable
as the single nanoantenna's length now is only 85 nm while in the
case of vacuum it was 130 nm.

With this extended study we see that in the case of vacuum  
the time dependence of the momentum absorption shows a beat pattern
due to two nearly identical resonance frequencies. Inside UDMA this
beating is not present and the absorption is less as it is shown in
{\color{black}
Fig. \ref{Fig-2}}
also.

Similar to 
\cite{PRX-E2022}
we also studied the time dependence of the momentum
fluctuation of the electrons.
Now the proton fluctuations were also studied. Initially the proton distribution slightly lags behind
the electrons indicating that the electrons are pulling
the protons. At later time as the laser drive is over
the two distributions become aligned in phase. This phenomena will need further investigation.

The time dependence of the energy absorption by a nanorod in 
UDMA polymer was also studied in the COMSOL Multiphysics model.
The results are similar, the main difference between the two models 
is arising from the different treatment of the conducting electrons. 
In the PIC model the conduction band electrons move freely and can escape 
leaving the gold ions behind, in the traditional, FEM approach the electrons' 
collective motion is taken into account indirectly through damping 
constants and they cannot leave from the surface of the nanorod.

\section{Acknowledgements}
Enlightening discussions with Johann Rafelski are gratefully acknowledged.
Horst St\"ocker acknowledges the Judah M. Eisenberg Professor Laureatus 
chair at Fachbereich Physik of Goethe Universit\"at Frankfurt.
We would like to thank 
the Wigner GPU Laboratory at the Wigner Research Center for Physics for 
providing support in computational resources.
This work is supported in part by
the Frankfurt Institute for Advanced Studies, Germany,
the E\"otv\"os Lor\'and Research Network of Hungary, 
the Research Council of Norway, grant no. 255253, and
the National Research, Development and Innovation Office of Hungary,
via the projects: Nanoplasmonic Laser Inertial Fusion Research Laboratory
(NKFIH-468-3/2021), Optimized nanoplasmonics (K116362), and
Ultrafast physical processes in atoms, molecules, nanostructures 
and biological systems (EFOP-3.6.2-16-2017-00005). L.P. Csernai acknowleges support from
Wigner RCP, Budapest (2022-2.2.1-NL-2022-00002)

\bigskip


\end{document}